# (N+2)-DIMENSIONAL ANISOTROPIC CHARGED FLUID SPHERES WITH PRESSURE: RICCATI EQUATION

**Naveen Bijalwan[1]**

General exact ($n+2$)-dimensional, $n \geq 2$ solutions in general theory of relativity of Einstein-Maxwell field equations for static anisotropic spherically symmetric distribution of charged fluid are expressed in terms of radial pressure. Subsequently, metrics ($e^\lambda$ and $e^\upsilon$), matter density and electric intensity are expressible in terms of pressure.
We extend the methodology used by Bijalwan (2011a, 2011c, 2011d) for charged and anisotropic fluid. Consequently, radial pressure is found to be an invertible arbitrary function of $\omega(=c_1+c_2r^2)$, where $c_1$ and $c_2(\neq 0)$ are arbitrary constants, and $r$ is the radius of star, i.e. $p_r = p_r(\omega)$. We present a general solution for static anisotropic charged pressure fluid in terms for $\omega$. We reduce to the problem of finding solutions to anisotropic charged fluid to that of finding solutions to a Riccati equation. Also, these solutions satisfy barotropic equation of state relating the radial pressure to the energy density.

## 1. Introduction

The problem of determination of exact solution of coupled Einstein-Maxwell equations for static spherical distribution of charged matter has attracted wide attention. These distributions constitute possible sources for Reissner-Nordstrom metric which uniquely describes the exterior field of spherically symmetric charged distribution of matter. Khadekar et al (2008) have generalized the technique used by Hajj-Boutros and Sfeila (1986) in the frame work of ($n+2$)-dimensional space-time and obtained exact solutions of Einstein-Maxwell field equations for static spherically symmetric distribution of charged perfect fluid. They have solved these field equations by using the equation of state $p = [\gamma - (n-1)]\rho$ and obtained the solutions for standard coordinates and isotropic coordinates system. Further, Harko (2000) worked on anisotropic charged fluid.

In this paper we extend methodology used by Bijalwan (2011a) to ($n+2$)-dimensional space to derive solutions of Einstein-Maxwell field equations for anisotropic charged fluid expressible terms of pressure. The pressure can be chosen arbitrarily such that solutions are comfortably matches with Reissner–Nordstrom solution at pressure-free interface yielding analytical expression for boundary radius $a$. In section 2 we derived the equation for anisotropic charged fluid.

## 2. Field Equations

Let us take the following spherically symmetric metric to describe the (n+2)-dimensional space-time of a static anisotropic charged fluid sphere

$$ds^2 = -e^{\lambda(r)}dr^2 - r^2 d\Omega^2 + e^{\upsilon(r)}dt^2. \qquad (2.1)$$

[1]FreeLancer, c/o Sh. Rajkumar Bijalwan, Nirmal Baag, Part A, Pashulock, Virbhadra, Rishikesh, Dehradun-249202 (Uttarakhand), India. ahcirpma@rediffmail.com

where $d\Omega^2 = d\theta_1^2 + \sin^2\theta_1 d\theta_2^2 + \sin^2\theta_1 \sin^2\theta_2 d\theta_2^2 + \ldots + \left[\prod_{i=1}^{n} \sin^2\theta_i\right] d\theta_n^2$,

and the coordinates are $x^1 = r, x^2 = \theta_1, x^3 = \theta_2 \ldots\ldots x^{n+1} = \theta_2, x^{n+2} = t$.

The field equations (using the geometric units $G = c = 1$) with respect to the metric (2.1) reduce to Harko(2000) for anisotropic fluid ($p_r \neq p_\perp$) and Khadekar et al (2008) for isotropic fluid ($p_r = p_\perp$)

$$\kappa T_1^1 = \frac{n\upsilon'}{2r}e^{-\lambda} + \frac{n(n-1)(e^{-\lambda}-1)}{2r^2} = 8\pi(p_r - k), \tag{2.2}$$

$$\kappa T_2^2 = \kappa T_3^3$$
$$= \left[\frac{\upsilon''}{2} - \frac{\lambda'\upsilon'}{4} + \frac{\upsilon'^2}{4} + \frac{(n-1)(\upsilon'-\lambda')}{2r} + \frac{(n-1)(n-2)}{2r^2}\right]e^{-\lambda} - \frac{(n-1)(n-2)}{2r^2}, \tag{2.3}$$
$$= 8\pi(p_\perp + k)$$

$$\kappa T_4^4 = \frac{n\lambda'}{2r}e^{-\lambda} + \frac{n(n-1)(1-e^{-\lambda})}{2r^2} = 8\pi(\rho + k), \tag{2.4}$$

where the prime (') denotes the differentiation with respect to $r$, where $p$ is fluid pressure and $\rho$ is the matter density and $k = -\frac{1}{8\pi}F_{(n+2)1}F^{(n+2)1}$, (2.5) represents the total charge contained with in the sphere of radius $r$.

Let us consider the barotropic equation of state $\kappa c^2 \rho = g(p_r)$. (2.8a)

On subtracting (2.2) from (2.4) gives

$$\left(\frac{\upsilon' + \lambda'}{r}\right)e^{-\lambda} = \kappa(c^2\rho + p_r) \tag{2.9a}$$

$$\left(\frac{\upsilon' + \lambda'}{r}\right)e^{-\lambda} = \kappa(c^2 g + p_r) \tag{2.9b}$$

Now, in order to solve (2.9b), let us further assume that metrics ($e^\lambda$ and $e^\upsilon$), and electric intensity are arbitrary functions of pressure $p(\omega)$ such that $\omega$ is some function of $r$ i.e.

$e^{-\lambda} = s(p_r(\omega)), e^\upsilon = h(p_r(\omega)), \qquad \kappa c^2 \rho = g(p_r(\omega))$ (2.8b)

On adding (2.2) and (2.4) gives

$$\frac{n}{2}\left(\frac{\upsilon' + \lambda'}{r}\right)e^{-\lambda} = 8\pi(\rho + p_r) \tag{2.9}$$

Substituting (2.8) in (2.9) leads to

$$\frac{n(\bar{\upsilon} + \bar{\lambda})}{2(c^2\rho + p)}e^{-\lambda}\frac{dp_r}{dr} = r \tag{2.10}$$

where overhead dash denotes derivative w.r.t. $p$ or $\omega$.

(2.10) yields
$p_r = f(c_1 + c_2 r^2) = f(\omega)$

i.e. function of '$c_1 + c_2 r^2$', where $c_1$ and $c_2 (\neq 0)$ are arbitrary constants.
such that

$$r = \sqrt{\frac{\omega - c_1}{c_2}}, \quad (\omega - c_1)c_2 > 0 \tag{2.11}$$

Assuming $f$ is invertible and $f^{-1}$ is inverse of $f$ then $f^{-1}(p_r) = \omega$.

Matter density, charge density and velocity of sound can be expressed using (2.8) in (2.10) and (2.2) as

$$8\pi\rho = nc_2\left(\frac{\bar{h}}{h}s - \bar{s}\right) - 8\pi \bar{p}_r \tag{2.12}$$

(2.12) is clearly a barotropic equation of state for it directly relates the radial pressure to the energy density.

$$8\pi k = c_2 \frac{(1-s)n(n-1)}{2(\omega - c_1)} - nc_2 s \frac{\bar{h}}{h} + 8\pi p_r \tag{2.13}$$

$$\sqrt{\frac{dp_r}{d\rho}} = 1 / \sqrt{\frac{nc_2}{8\pi}\left(\left(\frac{\bar{h}}{h}\right)s\right) - \bar{\bar{s}}) - \bar{p}_r} \tag{2.14}$$

Also, Equation (2.3) on using (2.12) and (2.13) leads to

$$\left[2n\frac{\bar{h}}{h} + 2(\omega - c_1)\left(\frac{\bar{\bar{h}}}{h}\right) + (\omega - c_1)\left(\frac{\bar{s}}{s}\right)\left(\frac{\bar{h}}{h}\right) + (\omega - c_1)\left(\frac{\bar{h}}{h}\right)^2 + (n-1)\frac{\bar{s}}{s} + \frac{(n-1)(n-2)}{2(\omega - c_1)}\right]s$$

$$-\frac{(1-s)(n-1)n}{2(\omega - c_1)} - \frac{(n-1)(n-2)}{2(\omega - c_1)} - \frac{8\pi(2p_r + \Delta)}{c_2} = 0 \tag{2.15}$$

where $\Delta = p_\perp - p_r$ is anisotropy factor.

$$\bar{\alpha} = -\frac{(2ns + (\omega - c_1)\bar{s})\alpha}{2(\omega - c_1)s} - \frac{\alpha^2}{2} + \eta(\omega) \tag{2.15a}$$

$$\eta(\omega) = -\frac{(n-1)\bar{s}}{2(\omega - c_1)s} - \frac{(n-1)(n-2)}{4(\omega - c_1)^2}$$

$$-\frac{(n-1)n}{4(\omega - c_1)^2} + \frac{1}{(\omega - c_1)s}\left(\frac{(n-1)n}{4(\omega - c_1)} + \frac{(n-1)(n-2)}{4(\omega - c_1)} + \frac{8\pi(2p_r + \Delta)}{c_2}\right)$$

Riccati equation (2.15a) can be solved for various values of $\eta(\omega)$.

If we now match the interior solution to the ($n+2$)-dimensional Reissner-Nordstrom exterior one at the boundary of star $a$, we obtain

$$e^{-\lambda}{}_{(r=a)} = e^{\nu}{}_{(r=a)} = 1 - \frac{2M}{a^{n-1}} + \frac{2q^2}{n(n-1)a^{2(n-1)}} = h(\omega(a)) = s(\omega(a)), \tag{2.16a}$$

where $q$ represents the total charge for $r = a$ and $M$ is total mass of the charged fluid sphere. Further, condition $p_{r(r=a)} = 0$ can be utilized to compute boundary radius $a$ of star.

## 3. Conclusions

We have studied the structure of the sources produced by ($n+2$)-dimensional Einstein-Maxwell field equations for anisotropic spherically symmetric distribution of charged pressure fluid in terms of radial pressure.